\documentclass[12pt]{article}
\usepackage{amstex}
\usepackage{amssymb}
\usepackage{epsfig}
\usepackage{cite}
\begin{document}
\null

\begin{center}
{\bf \large ${\cal O}(N_f\alpha^2)$ Corrections to Muon Decay}\\
\medskip
Robin G. Stuart\\
\medskip
{\sl Randall Laboratory of Physics\\
     Ann Arbor, Michigan 48109-1120, USA}
\end{center}

\begin{abstract}
The calculation of the ${\cal O}(N_f\alpha^2)$ corrections to muon decay
is described. These are the 2-loop diagrams containing a massless fermion
loop and they form an important gauge-invariant subclass. It is shown that
all such diagrams can be expressed in terms of a universal master integral.
We focus on the calculation of box diagrams and in particular on the removal
of their infrared divergences.
\end{abstract}

\section{Introduction}

Muon decay, $\mu^-\rightarrow\nu_\mu e^-\bar\nu_e$, has always
been a proving ground for both pure QED and electroweak radiative
corrections \cite{KinoSirlin,Sirlin80,Sirlin84}.
Until recently the muon decay coupling constant, $G_\mu$,
and the electromagnetic coupling constant,
$\alpha$, were by far the best measured electroweak observables and played
a pivotal r\^ole as input to the Standard Model. The accuracy of theoretical
predictions was limited by the errors on $M_Z$ which was taken as
the third input required to make the model predictive. Now the situation has
changed somewhat. Both $M_Z$ and $G_\mu$ are determined to an accuracy
of $2\times10^{-5}$ and it may be possible to reduce the error on $M_Z$
still further if the LEP beam energy calibration can be better understood.

By comparison electroweak calculations have attained an precision of only
a few permill. Obviously it would be desirable to have the results of full
2-loop electroweak calculations at our disposal and considerable progress
has been made in this direction \cite{Franzkowski}. Such calculations are
likely to be intractable when expressed in analytic form
and in practice one would quote only numerical results.
There do exist, however, a few 2-loop calculations for
which exact analytic results can be obtained in a compact closed form
\cite{Sirlin84,Frank,Barbieri,DennHollLamp,Fleischer}.
Here we discuss one such set of corrections. These are dubbed
${\cal O}(N_f\alpha^2)$ corrections to muon decay, where $N_f$ is
the number of light fermions.
Because $N_f$ is quite large, the ${\cal O}(N_f\alpha^2)$
corrections are expected to be a dominant subset of 2-loop graphs and
since $N_f$ provides a unique tag, the complete set of ${\cal O}(N_f\alpha^2)$
corrections contributing to a particular physical process will form a
gauge-invariant set. Diagrams of this type have been considered in
connection with the muon anomalous magnetic moment
\cite{CKM1}.

{\it A priori\/} the ${\cal O}(N_f\alpha^2)$ corrections can be expected to
contribute at the level of $1.5\times10^{-4}$. Without their calculation and
inclusion, theoretical predictions remain uncertain at least at this level.

If the fermions are assumed to be massless compared to $M_W$,
an excellent approximation for all but the third generation, then the
only `dimensionful' quantities appearing in the calculation are $M_W$,
$M_Z$ and $M_H$ and there are very few diagrams containing a Higgs.
The contributions from diagrams not involving Higgs bosons can then
only be a polynomial in $\sin^2\theta_W$ with coefficients involving
$\ln\cos^2\theta_W$, $\ln\pi$ and Euler's constant, $\gamma$.

Assuming massless fermions reduces the topologies of diagrams that
must be considered since the fermions can then
only couple to vector bosons but not to
Goldstones. However, the calculation is still a fully-fledged 2-loop
electroweak calculation requiring the complete renormalization of the
Standard Model at ${\cal O}(N_f\alpha^2)$. Although muon decay represents
a zero momentum transfer process, much can be learned from it about
the cancellation of divergences in high-energy processes. Indeed
unlike the case of the calculation of ${\cal O}(\alpha^2 m_t^4)$
corrections to the $\rho$-parameter \cite{Frank,Barbieri,Fleischer}, there
is a proliferation of diagrams involving counterterms and these
constitute significant fraction of the effort involved.

In addition to the calculation of diagrams at zero momentum transfer one
also requires the $W$ and $Z^0$ mass counterterms that must be obtained by
evaluating diagrams at high scales. In principle this can be done
in any renormalization scheme using methods expounded in
ref.s\cite{WeigleinScharfBohm,ScharfTausk} but the work is considerably
reduced by adopting the $\overline{\rm MS}$ scheme in which only the simpler
divergent parts of the diagrams are required.

In this talk we will summarize some of the salient features of the
calculation. In section 2 the master integral will be given and in section 3
we will concentrate on contributions coming from box diagrams. Diagrams of
this type do not occur in other 2-loop electroweak calculations performed to
date. It will be shown how the IR divergences can be
separated from these diagrams in a well-defined manner that
permits them to be meshed with bremsstrahlung diagrams.

The work reported here was undertaken with R.~Akhoury and P.~Malde. Full
details of the general methods used can be found in ref.\cite{AMS1} and
results for muon decay in ref.\cite{AMS2}.

\section{The Master Integral}

In what follows an anticommuting $\gamma_5$ is assumed.
A Euclidean metric with the square of time-like momenta being
negative will be used and all calculations will be done in
$R_{\xi=1}$ gauge.
The sine and cosine of the weak mixing angle,
$\theta_W$, will be denoted $s_\theta$ and $c_\theta$ respectively.
$\gamma_L$ and $\gamma_R$ are the usual left- and right-handed
helicity projection operators.

Many, but by no means all, ${\cal O}(N_f\alpha^2)$ diagrams are obtained
simply by inserting
a fermion loop into the boson propagator of a one-loop diagram.
Because the original one-loop diagram is often logarithmically
divergent, the fermion loop insertion will need to be calculated to
${\cal O}(n-4)$ in dimensional regularization, where $n$ is the dimension
of space-time. It turns out, however, that for all ${\cal O}(N_f\alpha^2)$
diagrams occurring in low-energy processes, it is possible to obtain
expressions that are exact in $n$.

For the massless fermion loop insertion, it may be shown that\\
\null\vskip -1.5cm
\vbox{
\begin{flushleft}
\epsfxsize 2.75 cm
\epsffile{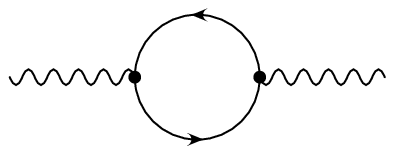}
\raisebox{2.35ex}[1.5 cm][0.5 cm]{$=$}
\end{flushleft}
\null\vskip -1.6cm\null
\begin{equation}
-\left(\delta_{\mu\nu}-\frac{\displaystyle p_\mu p_\nu}
                               {\displaystyle p^2}\right)
   (\beta_L\beta_L^\prime+\beta_R\beta_R^\prime)
   \frac{\displaystyle (n-2)}{\displaystyle (n-1)}
{\displaystyle \int}\frac{\displaystyle d^n q}
{\displaystyle i\pi^2}\frac{\displaystyle p^2}
{\displaystyle q^2(q+p)^2}
\end{equation}
}
where $\beta_L$, $\beta_R$ and $\beta_L^\prime$, $\beta_R^\prime$ of the
couplings of the attached vector bosons.

For processes at zero momentum transfer it is possible to
immediately reduce all tensor integrals that occur to scalar integrals
by using projection operator techniques \cite{AMS1}.
These scalar integrals can then be written as expressions involving
a general master integral,
\begin{equation}
\begin{split}
\int\frac{d^np}{i\pi^2}&\frac{1}{[p^2]^j[p^2+M^2]^k}
\int\frac{d^nq}{i\pi^2}\frac{1}{[q^2]^l[(q+p)^2]^m}\\
   &=\frac{\pi^{n-4}}{(M^2)^{k+j+l+m-n}}
          \Gamma\left(l+m-\frac{\textstyle n}{\textstyle 2}\right)
          \Gamma\left(\frac{\textstyle n}{\textstyle 2}-l\right)
          \Gamma\left(\frac{\textstyle n}{\textstyle 2}-m\right)\\
   &\times
    \frac{\Gamma(n-j-l-m)\Gamma(k+j+l+m-n)}
         {\Gamma\left(\frac{\textstyle n}{\textstyle 2}\right)
          \Gamma(k)\Gamma(l)\Gamma(m)\Gamma(n-l-m)}\\
\end{split}
\label{eq:MasterIntexpr}
\end{equation}

\section{Box Diagrams}

A useful set of identities for calculating one-loop box diagrams appears
in ref.\cite{SirlinBox}. They are, however, valid only for $n=4$
because of their intended use at one-loop. For general $n$ it may be shown
that these relations become
\begin{eqnarray}
{[}\gamma_\rho\gamma_\mu\gamma_\sigma\gamma_{L,R}{]}_1
{[}\gamma_\rho\gamma_\nu\gamma_\sigma\gamma_{L,R}{]}_2
&=&4\delta_{\mu\nu}{[}\gamma_\alpha\gamma_{L,R}{]}_1
                       {[}\gamma_\alpha\gamma_{L,R}{]}_2\nonumber\\
               &+&(n-4){[}\gamma_\mu\gamma_{L,R}{]}_1
                       {[}\gamma_\nu\gamma_{L,R}{]}_2\ \ \ \\
{[}\gamma_\rho\gamma_\mu\gamma_\sigma\gamma_{L,R}{]}_1
{[}\gamma_\rho\gamma_\nu\gamma_\sigma\gamma_{R,L}{]}_2
                   &=&4{[}\gamma_\nu\gamma_{L,R}{]}_1
                       {[}\gamma_\mu\gamma_{R,L}{]}_2\nonumber\\
               &+&(n-4){[}\gamma_\mu\gamma_{L,R}{]}_1
                       {[}\gamma_\nu\gamma_{R,L}{]}_2\ \ \ \\
{[}\gamma_\rho\gamma_\mu\gamma_\sigma\gamma_{L,R}{]}_1
{[}\gamma_\sigma\gamma_\nu\gamma_\rho\gamma_{L,R}{]}_2
                   &=&4{[}\gamma_\nu\gamma_{L,R}{]}_1
                       {[}\gamma_\mu\gamma_{L,R}{]}_2\nonumber\\
               &+&(n-4){[}\gamma_\mu\gamma_{L,R}{]}_1
                       {[}\gamma_\nu\gamma_{L,R}{]}_2\ \ \ \\
{[}\gamma_\rho\gamma_\mu\gamma_\sigma\gamma_{L,R}{]}_1
{[}\gamma_\sigma\gamma_\nu\gamma_\rho\gamma_{R,L}{]}_2
&=&4\delta_{\mu\nu}{[}\gamma_\alpha\gamma_{L,R}{]}_1
                       {[}\gamma_\alpha\gamma_{R,L}{]}_2\nonumber\\
               &+&(n-4){[}\gamma_\mu\gamma_{L,R}{]}_1
                       {[}\gamma_\nu\gamma_{R,L}{]}_2\ \ \ \ \
\end{eqnarray}
where the square brackets $[\ ]_1$ and $[\ ]_2$ indicate that the enclosed
$\gamma$-matrices are associated with the separate
external fermion currents $J_1$ and $J_2$ respectively.
These identities, along with projection operator techniques, may be used to
reduce box diagrams to the form $I\cdot{\cal M}_0$ where $I$ is a scalar
integral and ${\cal M}_0$ is the Born level matrix element.

All ${\cal O}(N_f\alpha^2)$ box diagrams correspond to the insertion of
a fermion loop into boson propagators in one-loop box diagrams.
The box diagrams, so obtained, are either logarithmically divergent or
have double poles at $n=4$ corresponding to mixed UV and IR divergences or
fermion mass singularities.

\begin{figure}[t]
    \begin{center}
    \epsfig{file=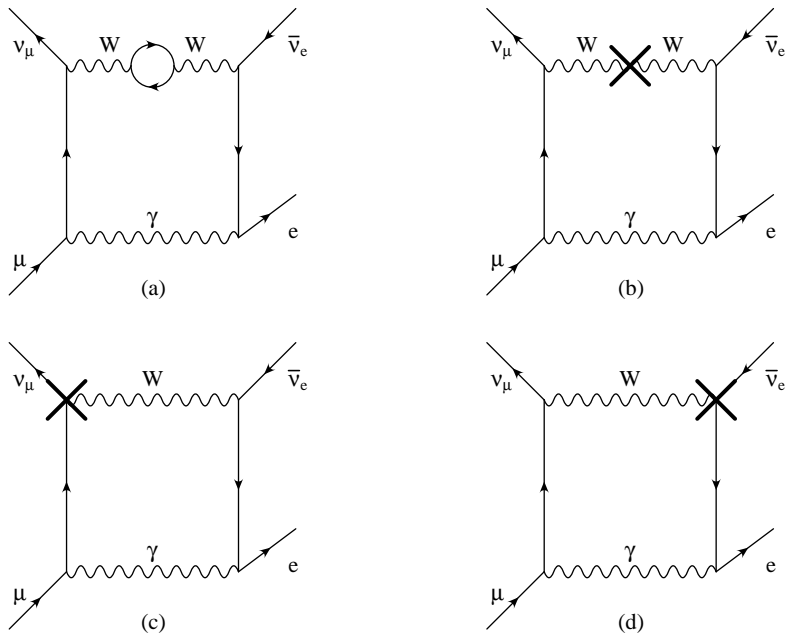}
    \caption{A class of box diagrams containing UV and IR divergences.}
    \end{center}
\end{figure}

Consider the diagrams of Fig.1. The crosses represent fermionic parts of
${\cal O}(\alpha)$ counterterms.
The fermion loop contribution in Fig.1a vanishes at $q^2=0$ where $q$ is
the 4-momentum in the photon propagator and the diagram is thus IR finite.
If the counterterms on the $W$ line in Fig.1b--d together vanish at
$q^2=0$ then they too will be IR finite however this depends on the
renormalization scheme that is chosen and is not generally the case.
It happens, for example, in the $\overline{\rm MS}$ but not in the
on-shell renormalization scheme.

In Fig.1b--d the insertion of counterterms
corresponds to replacing the $W$ propagator in the one-loop box diagram
\begin{eqnarray}
\frac{1}{q^2+M_W^2}&\rightarrow&
 \left(2\frac{\delta g}{g}\right)\frac{1}{q^2+M_W^2}
       -\frac{\delta M_W^2}{(q^2+M_W^2)^2}\\
 &=&\left(2\frac{\delta g}{g}\right)\frac{q^2}{(q^2+M_W^2)^2}
   -\frac{M_W^2}{(q^2+M_W^2)^2}
    \left\{\left(2\frac{\delta g}{g}\right)-\frac{\delta M_W^2}{M_W^2}\right\}
\nonumber\\
\label{eq:ctdecomp}
\end{eqnarray}
where $\delta g$ and
$\delta M_W^2$ are the $SU(2)$ coupling constant counterterm and $W$ mass
counterterm respectively.

The first term of (\ref{eq:ctdecomp}) yields an IR finite contribution
and its UV divergences
cancel against those of Fig.1a. The second term in (\ref{eq:ctdecomp})
is UV finite but generates an IR divergence and must be combined with
soft bremsstrahlung correction to produce a finite result. As mentioned
above the last term vanishes in the $\overline{\rm MS}$ scheme.
\begin{figure}[t]
    \begin{center}
    \epsfig{file=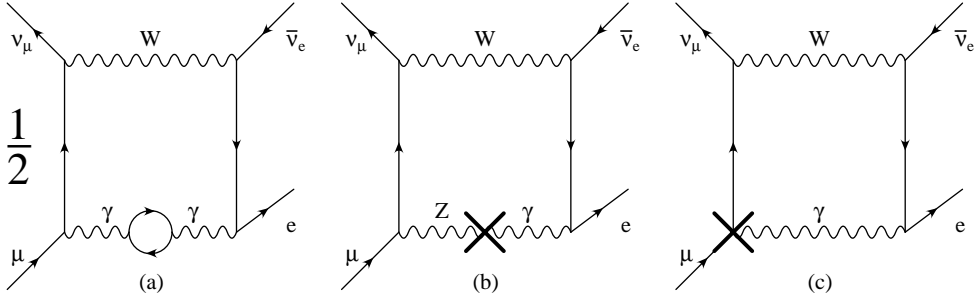}
    \caption{A class of box diagrams containing UV and IR divergences and
             fermion mass singularities.}
    \end{center}
\end{figure}

A similar thing happens for the diagrams shown in Fig.2 except that now
Fig.2a is both IR divergent and has a fermion mass singularity.
If we denote the fermion loop insertion in the photon
propagator as $(q^2\delta_{\mu\nu}-q_\mu q_\nu)\Pi_{\gamma\gamma}^\prime(q^2)$
then we may write
\begin{equation}
\Pi_{\gamma\gamma}^\prime(q^2)=
 [\widehat\Pi_{\gamma\gamma}^\prime(0)]
+[\Pi_{\gamma\gamma}^\prime(0)-\widehat\Pi_{\gamma\gamma}^\prime(0)]
+[\Pi_{\gamma\gamma}^\prime(q^2)-\Pi_{\gamma\gamma}^\prime(0)]
\label{eq:Pisplit}
\end{equation}
where
\begin{equation}
\widehat\Pi_{\gamma\gamma}^\prime(0)=s_\theta^2\frac{\delta g}{g}
                                    +c_\theta^2\frac{\delta g^\prime}{g^\prime}
\end{equation}
and $g^\prime$ is the $U(1)$ coupling constant.

The first term on the rhs of (\ref{eq:Pisplit}) yields an IR divergent
correction
that cancels against the IR divergences of the counterterm diagrams
Fig.2b and 2c leaving an overall UV divergence.
This remaining UV divergence cancels against those of other
box diagrams involving $Z^0$ bosons. The second term
in square brackets yields an IR divergent term that needs to be combined with
soft bremsstrahlung diagrams to yield a finite result. Finally the third term
gives a finite contribution that is singular for vanishing fermion masses.
It will be subject to a hadronic contribution and should therefore be evaluated
using dispersion relations. We do not consider the second and third term
further.

Once IR divergences are removed in this way, the ${\cal O}(N_f\alpha^2)$
corrections to the Born level matrix element, ${\cal M}_0$,
coming from box diagrams may be written as ${\cal M}_0\Delta r^{(2)}$ where
\begin{equation}
\begin{aligned}
\Delta r^{(2)}=
\left(\frac{g^2}{16\pi^2}\right)^2
\Bigg\{&\frac{2\ln^2c_\theta^2}{9 s_\theta^4}
    (8s_\theta^8+12s_\theta^6-131s_\theta^4+135s_\theta^2-45)\\
-&\frac{4}{3}\ln c_\theta^2(2s_\theta^2-1)
 -s_\theta^2[\Pi_{\gamma\gamma}^\prime(-M_Z^2)-\Pi_{\gamma\gamma}^\prime(0)]\\
+&\frac{\ln c_\theta^2}{s_\theta^2}
           [\Pi_{\gamma\gamma}^\prime(-M_Z^2)-\Pi_{\gamma\gamma}^\prime(0)]
    (2s_\theta^4-10s_\theta^2+5)
     \Bigg\}\ \ \ \ {}
\end{aligned}
\end{equation}
for each massless fermion generation.

This result is given in the on-shell renormalization scheme and the photon
vacuum polarization,
$\Pi_{\gamma\gamma}^\prime(-M_Z^2)-\Pi_{\gamma\gamma}^\prime(0)$, arises
from the definition of one-loop counterterms in this scheme. The
corresponding expression in the $\overline{\rm MS}$ scheme is somewhat
longer because of the presence of terms such as $\ln M_W^2/\mu^2$ and $\ln\pi$.


\begin{thebibliography}{99}

\bibitem{KinoSirlin} T. Kinoshita and A. Sirlin, {\sl Phys. Rev.}\
     {\bf 113} (1959) 1652.

\bibitem{Sirlin80} A. Sirlin, {\sl Phys.\ Rev.}\ {\bf D 22} (1980) 971.

\bibitem{Sirlin84} A. Sirlin, {\sl Phys.\ Rev.}\ {\bf D 29} (1984) 89.

\bibitem{Franzkowski} J. Franzkowski, these proceedings

\bibitem{Frank} J. J. van der Bij and F. Hoogeveen,
        {\sl Nucl.\ Phys.}\ {\bf B 283} (1987) 477.

\bibitem{Barbieri} R. Barbieri {\it et al.},
        {\sl Nucl.\ Phys.}\ {\bf B 409} (1993) 105.

\bibitem{DennHollLamp} A. Denner, W. Hollik and B. Lampe,
       {\sl Z. Phys.}\ {\bf C 60} (1993) 193.

\bibitem{Fleischer} J. Fleischer, O. V. Tarasov and F. Jegerlehner,
        {\sl Phys.\ Rev.}\ {\bf D 51} (1995) 3820.

\bibitem{CKM1} A. Czarnecki, B. Krause and W. Marciano,
        {\sl Phys.\ Rev.}\ {\bf D 52} (1995) 2619;
        A. Czarnecki, these proceedings

\bibitem{WeigleinScharfBohm} G. Weiglein, R. Scharf and M. B\"ohm,
        {\sl Nucl. Phys.}\ {\bf B 416} (1994) 606.

\bibitem{ScharfTausk} R. Scharf and J. B. Tausk,
        {\sl Nucl. Phys.}\ {\bf B 412} (1994) 523.

\bibitem{AMS1} R. Akhoury, P. Malde and R. G. Stuart, preprint UM-TH-96-16.

\bibitem{AMS2} R. Akhoury, P. Malde and R. G. Stuart, in preparation.

\bibitem{SirlinBox} A. Sirlin, {\sl Nucl.\ Phys.}\ {\bf B 192} (1981) 93.

\end{thebibliography}
\end{document}